\documentclass[conference]{IEEEtran}
\usepackage{float}
\usepackage{svg}
\usepackage{array}
\usepackage[T1]{fontenc}
\usepackage{textgreek}
\usepackage{makecell} 
\IEEEoverridecommandlockouts
\usepackage{lipsum}
\usepackage{flushend}
\usepackage{hyperref}
\hypersetup{colorlinks=true, linkcolor=blue, filecolor=magenta, urlcolor=cyan, citecolor=blue}
\usepackage{cite}
\usepackage{amsmath,amssymb,amsfonts}
\usepackage{algorithmic}
\usepackage{graphicx}
\usepackage{textcomp}
\usepackage{xcolor}
\usepackage{mathtools, nccmath}
\DeclarePairedDelimiter{\nint}\lfloor\rceil

\def\BibTeX{{\rm B\kern-.05em{\sc i\kern-.025em b}\kern-.08em
    T\kern-.1667em\lower.7ex\hbox{E}\kern-.125emX}}
\begin{document}

\title{A Practical Demonstration of DRL-Based Dynamic Resource Allocation xApp Using OpenAirInterface\\
%  {\footnotesize \textsuperscript{*}Note: Sub-titles are not captured in Xplore and
 }

\author{
\IEEEauthorblockN{Onur Sever\IEEEauthorrefmark{1}, Onur Salan\IEEEauthorrefmark{1}, Ibrahim Hokelek\IEEEauthorrefmark{1}, Ali Gorcin\IEEEauthorrefmark{1}\IEEEauthorrefmark{2} }
\IEEEauthorblockA{\IEEEauthorrefmark{1}Communications and Signal Processing Research (HISAR) Lab, TUBITAK BILGEM, Kocaeli, Turkiye}
\IEEEauthorblockA{\IEEEauthorrefmark{2}Faculty of Electrical and Electronics Engineering, Istanbul Technical University, Istanbul, Turkiye}
\IEEEauthorblockA{Email: \{onur.sever, onur.salan, ibrahim.hokelek, ali.gorcin\}@tubitak.gov.tr}
}

\maketitle

\begin{abstract}
Network slicing is a key enabler for providing a differentiated service support to heterogeneous use cases and applications in 5G and beyond networks through creating multiple logical slices. Resource allocation for satisfying diverse requirements of slices is a highly challenging task under time-varying traffic and wireless channel conditions. This paper presents a deep reinforcement learning (DRL) approach for allocating radio resources to slices, where the objective is to meet the latency requirement of the low-latency slice without jeopardizing the performance of the other slice. The proposed DRL approach is implemented within an open source mobile network emulator, namely OpenAirInterface, to create an O-RAN compliant end-to-end 5G network capable of dynamic resource allocation capabilities. The intelligent resource allocation mechanism operates on the RAN Intelligent Controller (RIC) as an xApp, enabling monitoring and dynamic resource control of the gNB through the E2 interface. The results demonstrate that the latency requirement of the low-latency slice is met under extremely loaded traffic scenarios, where the trained DRL model deployed on the near-RT RIC platform is used to dynamically allocate the radio resources to the slices.

\end{abstract}


\section{Introduction}
The disaggregation of the RAN into Centralized Unit (CU), Distributed Unit (DU), and Radio Unit (RU) by 3GPP is an important concept for enabling a modular and flexible system architecture. Open Radio Access Network (O-RAN) extends this functional split by promoting software-defined RAN intelligent controller (RIC) frameworks through standardised open interfaces for service management and orchestration (SMO) \cite{polese2023understanding}. Artificial intelligence (AI) and machine learning (ML) empowered by big data, cloud and edge networking, and virtualization are other enabling concepts from the system perspective for designing flexible and programmable network infrastructure. These concepts will enable zero-touch network management while significantly lowering CAPEX and OPEX in the longer term, and open architectures are anticipated to be the main driver behind this innovation. It is a highly challenging task to develop AI/ML based intelligent applications which can automate network functions as the network conditions change dynamically due to mobility and unprecedented traffic variations. 

An intelligent SMO is crucial for not only optimizing the overall network performance but also satisfying Quality of Service (QoS) requirements of diverse services and use cases. A related concept is network slicing (NS) which divides a single physical network into multiple logical slices to meet specific application requirements of 5G and beyond networks. NS can be categorized into core network (CN) slicing \cite{blenk2015survey} and RAN slicing \cite{ksentini2017toward}. The latter remains a particularly challenging and interesting problem due to time-varying traffic and dynamic wireless channel conditions \cite{zangooei2023reinforcement}.

In the O-RAN standards, the non-real time RIC (non-RT RIC) operates on a time scale longer than one second, with the main goal of supporting long-term radio resource management and policy optimization through rApps with AI and ML models. Similarly, the near-real time RIC (near-RT RIC) operates on a time scale of 10 ms to 1 second, managing and controlling valuable radio resources on the RAN side through xApps \cite{RIC}. There are numerous studies exploring the integration of RIC with advanced 6G RAN technologies, including Reconfigurable Intelligent Surfaces (RIS), MIMO \cite{ric-mimo}, and Network Slicing \cite{dai2024ran}. In this context, reinforcement learning (RL) becomes a prominent tool to optimize resource allocation and decision-making processes in highly dynamic network conditions, such as varying channel capacities, traffic loads, and interference levels \cite{cui2022qos,akyildiz2024hierarchical,alsenwi2021intelligent,zhou2021ran}. In these studies, the action space generally represents the allocation of radio resources, including frequency blocks and transmission power, to users. This approach allows for continuous adaptation and optimization of network policies, enabling more efficient automation of network functions.

\begin{figure*}[ht]
    \centering
    \includegraphics[width=0.95\textwidth]{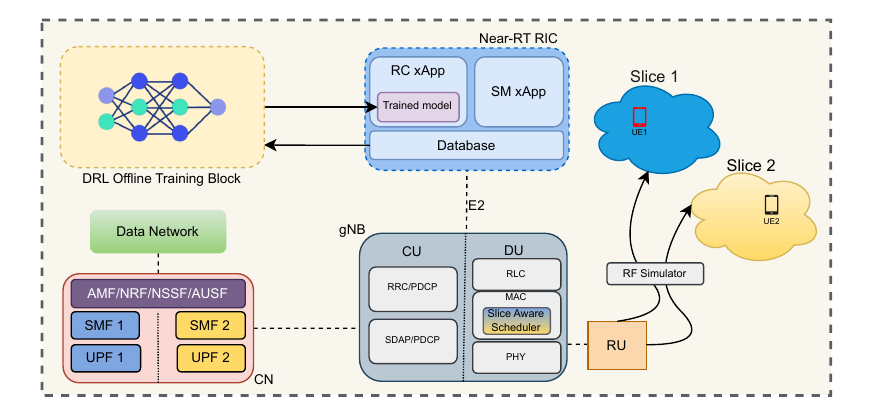}
    \caption{The convergence of the DRL algoritm}
    \label{fig:avgreward}
\end{figure*}

In this paper, we propose a DRL-based intelligent resource scheduling application (xApp) on the near-RT RIC within the Open RAN (O-RAN) to autonomously allocate the limited radio resources to network slices. The state space includes the traffic arrival rates of the network slices while the action space includes the allocation ratio of the avialable resource blocks to the low latency slice.  The reward function is designed in such a way that the minimal allocation ratio that satisfy the delay requirement of the low latency slice is utilized. We implemented the proposed DRL approach using open source 5G mobile network emulator, namely OpenAirInterface (OAI) which provides a complete 5G protocol stack for user equipment (UE), gNodeB (gNB), and the core network (CN) \cite{nikaein2014openairinterface}. The DRL models are first trained by generating a comprehensive traffic data set using the iperf tool within the OAI based end-to-end 5G network emulator, where real-time traffic ranging from 10 Mbps to 140 Mbps is generated for the low-latency slice while the traffic rate for the other slice is generated with a constant rate of around 117 Mbps. The DRL-based resource allocation mechanism with emphasis on the self-organizing network (SON) aspect is an important enabler  to provide robustness and elasticity for mobile network operators. 

The rest of the paper is organized as follows. Section \ref{sectionIII} introduces the system model of the DRL-based resource allocation mechanism for network slicing capable O-RAN networks. Section \ref{sectionIV} describes the proposed DRL-based resource allocation algorithm. Section \ref{sectionV} presents the experiment results using open source 5G emulators including CN, gNB, UEs, and RIC. Section \ref{sectionVI} concludes the paper and highlights the future work plan.

\section{SYSTEM MODEL}
\label{sectionIII}

Fig. \ref{fig:sysmod} illustrates the system model of the DRL-based intelligent resource allocation mechanism in an O-RAN complaint mobile cellular network consisting of a CN, a gNB, a near-RT RIC, and two UEs. The CN provides a Data Network access by creating two independent slices (Slice 1 in blue, Slice 2 in yellow) with a differentiated service support. A dedicated user plane function (UPF) and session management function (SMF) pair is created for each slice within the CN. Each NS is distinguished from each other using a single network slice selection assistance information (S-NSSAI), which obtains a slice/service type (SST) and a slice differentiator (SD). SST defines the characteristics of an NS using an 8-bit value, similarly, the optional SD distinguishes slices of the same type using a 24-bit value. A slice aware gNB connected to the CN provides a radio access functionality to the users of two slices. The near-RT RIC platform hosts Resource Control (RC) and Slice Monitoring (SM) xApps, which communicate with the gNB through the E2 interface. The SM xApp continuously collects key performance measurements (KPMs) from the gNB while the RC xApp is responsible for allocating the radio resources to the slices.

First of all, the data sets consisting of different traffic arrival rates and modulation and coding schemes (MCSs) of the users are generated using the end-to-end 5G emulator. The SM xApp collects these data sets and writes them to the database. The DRL Offline Training Block utilizes this dataset to obtain the trained DRL model which includes the resource allocation weights of the slices (i.e., DRL actions) corresponding to different traffic arrival rates of the slices (i.e., DRL states). The RC xApp is responsible for intelligently allocating the radio resources to the slices using the trained DRL model. During the DRL testing phase, the SM xApp provides the state information to the RC xApp by continuously monitoring the gNB performance statistics and the RC xApp determines the resource allocation weights of the slices (the DRL actions) from the trained DRL model. Once these weights are provided to the gNB, the scheduler provided by the OAI emulator ensures that the resources are allocated according to the RC xApp provided resource allocation weights. When the trained DRL model is not used, the default resource allocation weights are set to equal in the OAI emulator.  

\section{DRL-BASED RESOURCE ALLOCATION} 
\label{sectionIV}
An agent in a typical RL mechanism interacts with its environment to maximize its rewards. One technique for predicting the reward of an action is Q-learning which is iteratively updated to obtain the \textbf{Q(s,a)} function. A deep neural network (DNN) is utilized to approximate the Q-function in the deep Q-Network (DQN) which uses a $\epsilon$-greedy policy to choose actions. Experiences are randomly selected from the replay memory to avoid biases in learning. The aim of training is to decrease the discrepancy between the target and predicted Q-values using a loss function \cite{akyildiz2024hierarchical}.

For the sake of simplicity, this paper presents the DRL-based resource allocation when there are two diverse slices in the network. However, the DRL approach can be further extended to support more than two slices. The DRL policy for the first slice is to use the minimum number of resource blocks (RBs) that can satisfy the pre-defined delay threshold ($D_T$) while the policy for the second slice is to utilize the remaining RBs to maximize its throughput. The DQN algorithm dynamically modifies the resource allocation weights of the slices to ensure that each slice adheres to its pre-defined policies and maximizes the overall network performance.

\subsection {State}

Our state space is defined by all possible traffic arrival rates of the selected slice. The state space of the model would become extremely large if the traffic arrival rate of the slice were used without a quantization process. To reduce the state space, the quantized traffic arrival rate ($\hat{t}{_\text{arr}}$) is calculated as follows:

\begin{equation}\label{eq:quantization func}
\hat{t}{_\text{arr}} = \nint[\Big]{\frac{t_\text{arr} - L}{s}}s + L
\end{equation}
where $t_\text{arr}$, $s$, and $L$ represent traffic arrival rate,   quantization step, and the minimum traffic arrival rate, respectively. If $\hat{t}{_\text{arr}}$ exceeds the maximum traffic arrival rate ($H$), the clipped traffic arrival rate is calculated using the following formula:
\begin{equation}
    \bar{t}{_\text{arr}}= 
\begin{dcases}
    H,& \text{if  } \hat{t}{_\text{arr}}\geq H\\
    L,& \text{if  } \hat{t}{_\text{arr}}\leq L\\
    \hat{t}{_\text{arr}},              & \text{otherwise}
\end{dcases}
\end{equation} 

\subsection {Action}
The action space of the DRL includes the resource allocation weight of the selected slice as follows: 

\begin{equation}
    A(s) = \{ a_{1}, a_{2}, \dots, a_{n} \}
\end{equation}
where $0 < a_{\text{i}} < 100 $. Here, the resource allocation weight of the selected slice represents the percentage of RBs allocated to the selected slice. In this study, the resource allocation weight is changed from $10\%$ to $90\%$ by the increment of $5\%$.

\subsection {Reward}

The DRL reward function is designed as follows:

\begin{equation}
    A^*(s) = \{a_{k} \in A(s): UE_{d} < D_{T} \} 
\end{equation}
where $UE_{d}$ represents the delay in the service data unit (SDU). 

\begin{equation}
R = -1 \cdot \lvert a_i - \min \{ A^*(s) \} \rvert
\end{equation}
%$\lvert a_i - \min \{ A^*(s) \} \rvert$ 
The reward function selects an action which provides the minimum delay requirement of the selected slice using the minimum number of RBs. As the absolute difference between the action $a_i$ selected by the agent and  $ \min \{ A^*(s) \}$ increases, the agent moves further away from the optimal action. Conversely, as the absolute difference decreases, the agent approaches the optimal action.

\begin{table}[h!]
\centering
\small
\scriptsize % Yazı boyutunu küçültür
\setlength{\tabcolsep}{3pt} % Sütunlar arasındaki boşluğu daraltır
\begin{tabular}{|p{3cm}|p{3cm}|}
\hline
\textbf{Parameters} & \textbf{Value} \\ \hline
%Fully connected 5 layers & \makecell{$||S|| \times 256$ \\ $256 \times 256$ \\ $\vdots$ \\ $256 \times |A(s)|$} \\ \hline
%State Space & $||S||$ \\ \hline
%Action Space & \texttt{|A|} \\ \hline
%Hidden Layer Size & 256 \\ \hline
Model Architecture & Fully connected 5 layers \\ \hline
Hidden Layer Size & 256 \\ \hline
Activation Function & ReLU \\ \hline
Adam optimizer & $\beta_{1}=0.9$, $\beta_{2}=0.999$  \\ \hline
Learning Rate & 0.001 \\ \hline
Batch Size & 128 \\ \hline
Replay Buffer Size & 10,000\\ \hline
Target Network Update & Every \textit{N} episodes \\ \hline
Discount Factor ($\gamma$) & 0.99 \\ \hline
Exploration Strategy & $\varepsilon$-greedy (1 $\rightarrow$ 0.01) \\ \hline
Training Episodes & 1500 \\ \hline
Maximum step number & 100 \\ \hline
\end{tabular}
\vspace{0.2cm}
\caption{Training Model Parameters}
\label{tab:dqn_parameters}
\end{table}

\section{PERFORMANCE EVALUATION}
\label{sectionV}

In this section, we present the experimental setup and measurement results using the O-RAN compliant end-to-end 5G network emulator including 5G CN, 5G gNB, and UEs developed by OpenAirInterface (OAI). FlexRIC developed by the MOSAIC5G project group is utilized as the near-RT RIC platform \cite{schmidt2021flexric}. The OAI gNB emulator with 106 PRBs and 30 kHz SCS runs on a PC (Ubuntu 20.04.6 LTS, Intel i9-13900K), together with two OAI UEs and FlexRIC platform. The OAI CN emulator runs on a separate PC (Ubuntu 22.04.5 LTS, Intel i7-1260P). For all experiments, the RF simulator is utilized to simulate wireless channels between the gNB and UEs.

In this study, the training of the DRL agent is performed using the DQN algorithm using the parameters in Table~\ref{tab:dqn_parameters}. For generating the data set using the iperf tool, the traffic arrival rate is varied from 10 Mbps to 140 Mbps for the selected slice while the traffic arrival rate of the other slice is kept constant around 117 Mbps. The SM xApp periodically collects performance statistics including downlink traffic volume and latency from the gNB. In each traffic arrival scenario, the SM xApp writes the collected performance statistics to the database at 100 ms intervals for a duration of two minutes. The DRL agent is trained using the datasets acquired from the database. Fig. [\ref{fig:avgreward}] shows that the average reward increases with the number of episodes, demonstrating that the DRL algorithm successfully converges.

\begin{figure}[h]
    \centering
    \includegraphics[width=0.55\textwidth]{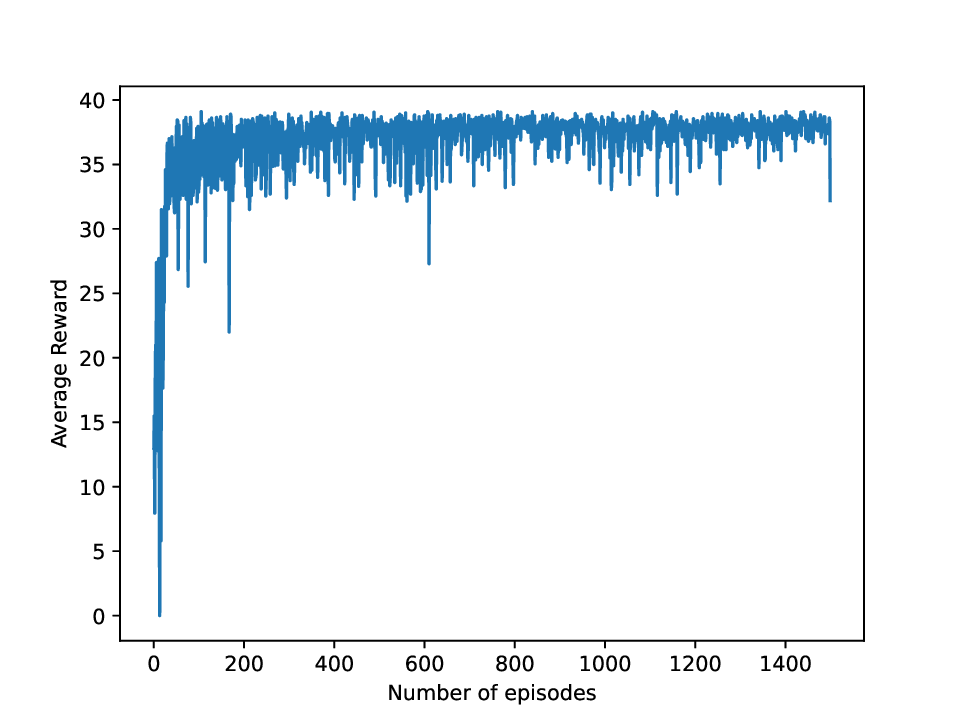}
    \caption{The convergence of the DRL algoritm}
    \label{fig:avgreward}
\end{figure}

\begin{figure}[h]
    \centering
    \includegraphics[width=0.55\textwidth]{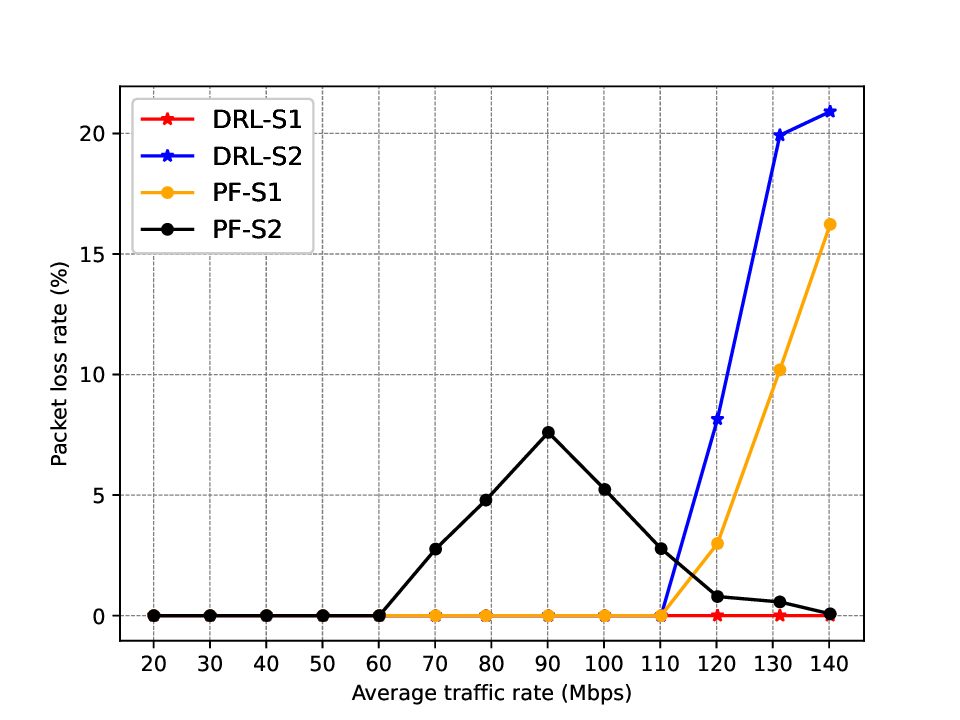}
    \caption{The packet loss rate results for the DRL and PF approaches}
    \label{fig:dropratio}
\end{figure}

\begin{figure}[h]
    \centering
    \includegraphics[width=0.55\textwidth]{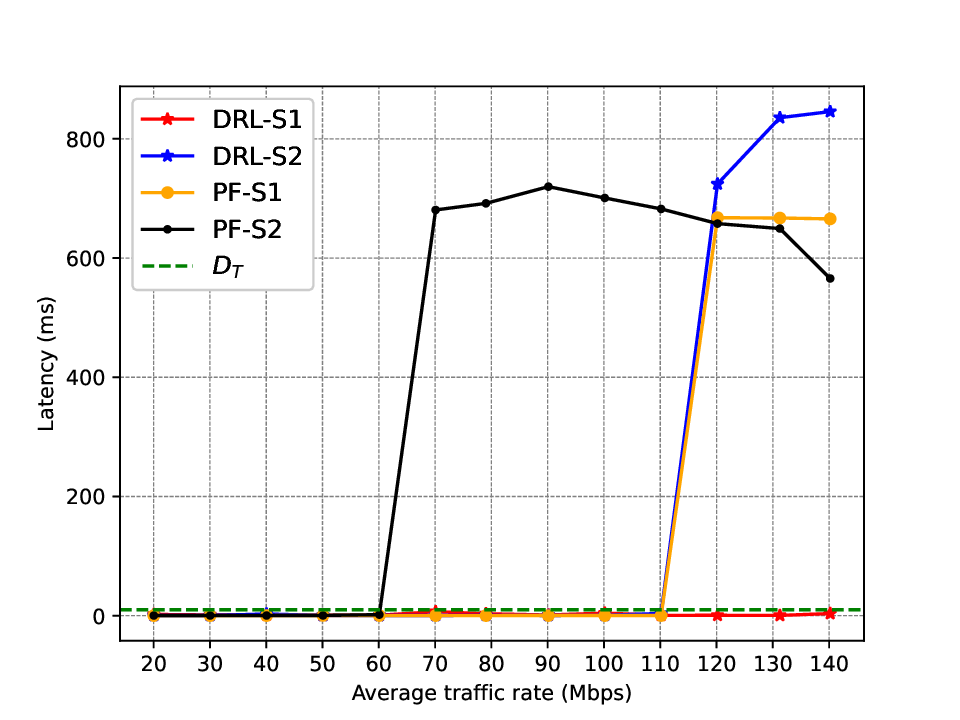}
    \caption{The latency results for the DRL and PF approaches}
    \label{fig:latency}
\end{figure}

\begin{figure}[h]
    \centering
    \includegraphics[width=0.55\textwidth]{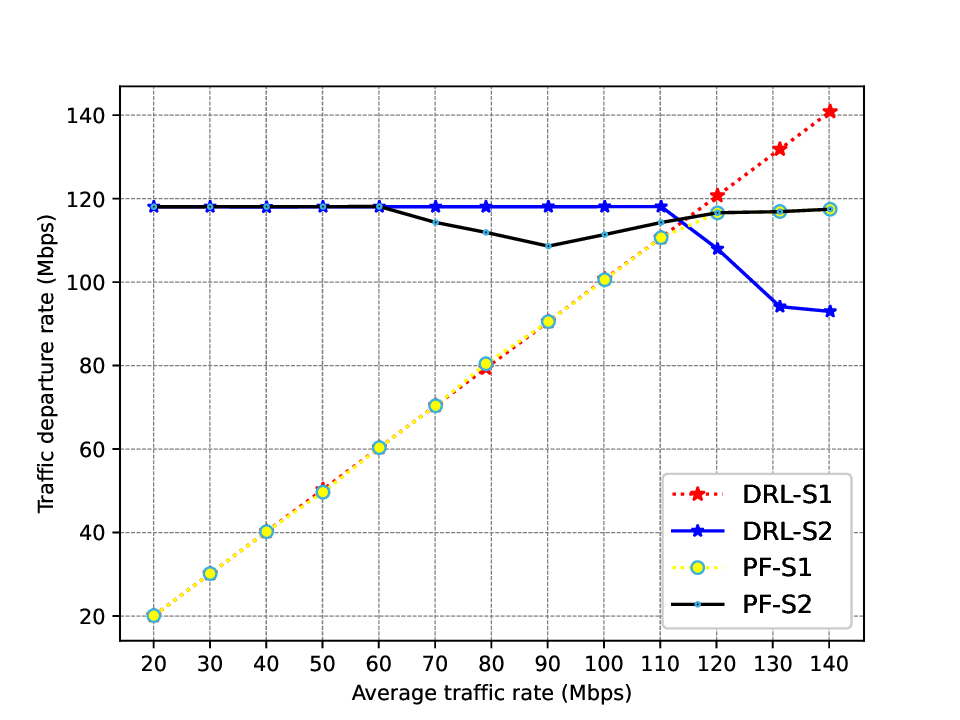}
    \caption{The traffic departure rate results for the DRL and PF approaches}
    \label{fig:DLrate}
\end{figure}

The performance of the proposed DRL approach is compared with the proportional fair (PF) scheduling algorithm which is available in the OAI emulator. Note that we conducted 13 different experiments corresponding to 13 different traffic arrival rate combinations of the slices for each algorithm as depicted in Figs. \ref{fig:dropratio}-\ref{fig:DLrate}, where the x-axis represents the average traffic arrival rate of Slice 1. For example, the arrival rate of Slice 2 is set to 117 Mbps for 13 experiments while the arrival rate of Slice 1 is increased from 20 Mbps to 140 Mbps by the increment of 10 Mbps. For each experiment with the duration of 5 minutes, the average performance results (packet loss rate, latency, traffic departure rate) are reported. %In the experiments, the DRL approach makes a new resource allocation weight decision in every 10 seconds.

Fig. \ref{fig:dropratio} illustrates the average packet loss rate results of the DRL and the PF approaches for each slice. The DRL method achieves zero packet loss for Slice 1 (DRL-S1) under any traffic arrival rate scenario as the policy specifies the reliability objective for Slice 1. When the traffic arrival rate of Slice 1 increases beyond 60 Mbps, we observed that there are non-zero packet loss rates for Slice 2 while the DRL provides no packet loss for both slices up to the traffic arrival rate of 110 Mbps. When the traffic arrival rate of Slice 1 goes beyond 110 Mbps, the DRL still maintains zero packet loss at the expense of higher packet loss rates for Slice 2. This is an expected result according to the network administrator specified policy. For the PF approach, the traffic arrival rate of Slice 1 is beyond 120 Mbps, the packet loss rates of Slice 2 is lower than Slice 1. This might be due to the fact that the PF provides equal RB allocation for each slice under heavy traffic loads while the arrival rate of Slice 2 is lower.

Fig. \ref{fig:latency} shows the average latency results of the DRL and the PF approaches for all slices. The DRL method successfully satisfies the latency requirement of Slice 1 (DRL-S1) under any traffic arrival rate scenario by keeping the average latency results of Slice 1 below the pre-defined threshold value. When the traffic arrival rate of Slice 1 increases beyond 60 Mbps, we observed that the latency of PF-S2 begins to rise sharply while both DRL-S1 and DRL-S2 result in low latency values below the threshold up to the traffic arrival rate of 110 Mbps for Slice 1. Using the same traffic arrival rates, Fig. \ref{fig:DLrate} illustrates the average traffic departure rate results of the DRL and the PF approaches at the gNB for both slices. The results are consistent with the packet loss rate and latency resuts in Fig. \ref{fig:dropratio} and Fig. \ref{fig:latency}, respectively.

\section{CONCLUSION} 
\label{sectionVI}
This paper proposes a DRL-based intelligent resource allocation mechanism which is implemented within an open source mobile
network emulator. We performed a sucessful demonstration of the DRL-approach using O-RAN compliant end-to-end 5G network, where the dynamic resource allocation is performed by an xApp running on the near-RT RIC platform. The experiment results demonstrate that the DRL approach satisfies the latency requirement of the selected slice under heavily loaded traffic scenarios. The performance results show that the DRL-based RAN slicing approach optimizes the resource utilization compared to the default resource allocation in the OAI platform. For future work, we will validate the proposed method in an over-the-air environment instead of using an RF simulator. 

\bibliographystyle{IEEEtran}
\bibliography{conference.bib}

\vspace{12pt}

\end{document}